\documentclass[10pt]{iopart}
\usepackage{epsfig}
\usepackage{iopams}
\usepackage{mathptmx}
\usepackage{eucal}
\usepackage{bm}

\newcommand{\re}{\,\mathrm{Re}\,}
\newcommand{\im}{\,\mathrm{Im}\,}
\newcommand{\const}{\,\mathrm{const}\,}
\newcommand{\W}{\widetilde{W}}
\newcommand{\go}{\bar{g}}
\newcommand{\ro}{\bar{\rho}}
\begin{document}

\title{Nonequilibrium and relaxation effects in tunnel superconducting junctions}

\author{E V Bezuglyi$^{1,2}$, A S Vasenko${^3}$ and E N Bratus'$^1$}
\address{$^1$ B Verkin Institute for Low Temperature Physics and Engineering, 61103 Kharkov, Ukraine}
\address{$^2$ Chalmers University of Technology, S-41296 G\"oteborg, Sweden}

\address{$^3$ National Research University Higher School of Economics, 101000 Moscow, Russia}

\ead{eugene.bezuglyi@gmail.com}

\begin{abstract}
The specific property of a planar tunnel junction with thin-film diffusive plates and long enough leads is an essential enhancement of its transmission coefficient compared to the bare transparency of the tunnel barrier \cite{BezuglyRC06,BezuglySUST}. In voltage-biased junctions, this creates favourable conditions for strong nonequilibrium of quasiparticles in the junction plates and leads, produced by multiparticle tunneling. We study theoretically the interplay between the nonequilibrium and relaxation processes in such junctions and found that nonequilibrium in the leads noticeably modifies the current-voltage characteristic at $eV > 2\Delta$, especially the excess current, whereas strong diffusive relaxation restores the result of the classical tunnel model. At $eV \leq 2\Delta$, the diffusive relaxation decreases the peaks of the multiparticle currents. The inelastic relaxation in the junction plates essentially suppresses the $n$-particle currents ($n>2$) by the factor $n$ for odd and $n/2$ for even $n$. The results may be important for the problem of decoherence in Josephson-junction based superconducting qubits.
\end{abstract}


\vspace{2pc}

\ioptwocol

\section{Introduction}

\begin{figure}[tb]
\begin{center}
\epsfxsize=8cm\epsffile{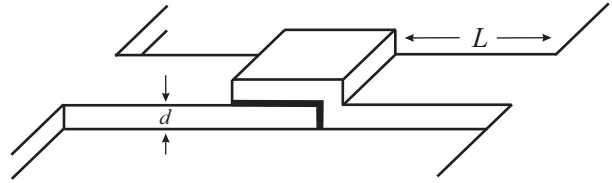}
\end{center}
\caption{The model of a planar SIS junction. The junction with upper and lower plates of thickness $d \ll \xi_0$ is connected to the bulk superconducting electrodes by leads of the length $L \gg \xi_0$.}
\label{model}
\end{figure}

Mesoscopic-size superconducting tunnel structures have become increasingly important devices in applications ranging from medical and astrophysical sensors to quantum computing due to their minimal dissipation at low
temperatures. This dissipation is often parameterized by the subgap conductance in parallel with an ideal tunnel element. The reason for this conductance is the quasiparticle current at voltages smaller than the superconducting gap, $eV < \Delta$. The main mechanism of charge transport in these conditions is the multiparticle tunneling (MPT) \cite{MPT} or, equivalently, coherent multiple Andreev reflections (MAR) of quasiparticles from the superconducting electrodes \cite{KBT}. In experiments, this process manifests itself by current steps at the voltages $eV = 2\Delta/n$ ($n = 1,2,\ldots$), which form the subharmonic gap structure (SGS) of the current-voltage characteristics (CVC). In the ballistic regime, the relation between the heights of the CVC consecutive steps is $D/2$, where $D$ is the transparency coefficient
of the tunnel barrier \cite{Bratus95,Averin95,Cuevas96}. This relation is performed relatively well for the point atomic-size contacts \cite{Jan2000}, but in mesoscopic tunnel junctions (see Figure \ref{model}) it turns out to be much larger which results in abnormally large subgap conductance \cite{Gubrud2001} and thus excess dissipation. The latter may be a source of the energy relaxation in superconducting qubits \cite{Makhlin,Wendin,Paauw,Martinis} and in tunable resonators \cite{Sandberg,Harvey,Levenson-Falk}. In single-electron turnstiles, this leakage may limit the ultimate accuracy of a future current standard \cite{Pekola1}. It can also serve as a limitation of the performance of microcoolers, based on hybrid superconducting tunnel structures \cite{Muhonen,Rajauria,VH,VBCH,Kawabata}. It is therefore important to study the mechanisms of subgap electron transport in mesoscopic-size superconducting tunnel structures.

For a long time, the common explanation for the enhanced SGS in mesoscopic junctions was the reference to the imperfect tunnel barrier or to
possible presence of the resonant levels or pinholes providing increased transparency \cite{KBT, Marcus, Kle}. This explanation was indeed found to be valid for high-transmissive junctions \cite{Naveh}. However, it is not the case for low-transmissive junctions with good insulating layers. Recent experiments of Greibe et al. \cite{Greibe} on Al/AlO$_\mathrm{x}$/Al junctions rule out pinholes as the origin of the excess current. In our previous papers \cite{BezuglyRC06, BezuglySUST} we have suggested a mechanisms of abnormally large subgap current, alternative to the ``pinhole'' explanation. It takes in consideration the effect of scattering of charge carriers in the diffusive banks of the junction, which results in an effective increase of the tunnel barrier transparency. The resulting physical picture is as follows: the tunneling processes induce the local nonzero density of states inside the bulk energy gap in the vicinity of the tunnel junction. This allows
quasiparticles to overcome the energy gap at $eV < 2\Delta$ in several steps, by repeated bouncing between the junction electrodes, i.e. by MAR
processes. The subgap current is calculated by considering an equivalent ``electrical network'' representing the tunneling process. However, we have neglected the quasiparticles nonequilibrium in the junction leads, as well as the energy relaxation of the subgap quasiparticles which may essentially modify the results of \cite{BezuglyRC06,BezuglySUST}. In this paper we study different aspects of such nonequilibrium and relaxation effects.

The paper is organized as follows. We start with the description of our junction model, basic equations and adopted approximations in section \ref{SecModel}, which is basically the summary of our papers \cite{BezuglyRC06,BezuglySUST}.
In section \ref{SecNoneq} we study the effect of strong nonequilibrium of quasiparticles in the leads, produced by MPT; this section includes also the calculation of the excess current and the peaks of multiparticle currents. At $eV > \Delta$, strong diffusive relaxation restores the results of the classical tunnel model for the excess current and the CVC shape; at $eV \leq 2\Delta$, it decreases the peak values of the multiparticle currents. The effect of the subgap quasiparticle relaxation, which essentially suppresses the $n$-particle currents by the factor $n$ for odd and $n/2$ for even $n$, is evaluated in section \ref{SecRelax}. We summarize the results in section \ref{SecSummary}.

%

\section{Model and basic equations}\label{SecModel}

In this paper we consider a planar superconductor-insulator-superconductor (SIS) junction sketched in Figure \ref{model}. It consists of an insulating layer (shown with the bold black line) with the transparency $D$ attached to bulk superconducting electrodes via two superconducting leads of the length $L$ and thickness $d$ (this is a typical configuration common to many physical applications). We emphasize that such situation is very different from the point-contact structure in which the tunnel barrier is directly connected to the massive equilibrium electrodes (reservoirs); in our case, the barrier is surrounded by the segments of the thin-film superconducting electrodes (referred to below as ``plates'', similarly to the capacitor plates), in which both the spectral characteristics and quasiparticle population may be far from their equilibrium values.

We consider a diffusive limit, in which the elastic scattering length $\ell$ is much smaller than the coherence length $\xi_0 = \sqrt{\mathcal{D}/2\Delta}$, where $\mathcal D$ is the quasiparticle diffusion coefficient (we assume $\hbar = k_{\rm B} = 1$). We assume the thickness $d$ to be much smaller than the Josephson penetration depth which implies homogeneity of the current along the junction, and the length $L$ of the leads to be much larger than $\xi_0$ but smaller than the inelastic scattering length $\ell_\epsilon$ (in the opposite case, $L > \ell_\epsilon$, the latter will qualitatively play the role of $L$ in the equations below). Under these conditions, it is possible to reduce the electron transport equations in this essentially 2D case to the 1D problem by formulating effective boundary conditions at the junction following the method suggested by Volkov \cite{VolkovSIN} and used in our previous works \cite{BezuglyRC06, BezuglySUST}. The planar SIS junctions were also considered in \cite{Berthod}.

In this section we briefly review the key points of our approach \cite{BezuglyRC06, BezuglySUST} used in this paper. The theory is based on the equation of nonequilibrium superconductivity, $[\check{H}, \check{G}] = \rmi \mathcal{D} \nabla [ \check{G} \nabla\check{G}  ] $, for the $4\times 4$ matrix two-time Green's function $\check{G}({\bf r}, t_1, t_2)$ in the diffusive environment of the barrier described by the Hamiltonian $\check{H}$ \cite{LOnoneq, Belzig}, with the boundary conditions of local equilibrium in bulk superconducting electrodes far from the contact. Analytical solutions of this equation can be constructed in the adiabatic limit of small applied voltage $eV \ll \Delta$ \cite{Bezugly2005}. At larger voltages $eV \sim \Delta$, due to complicated mathematical structure of this equation, its solution can be obtained only by means of numerical or approximate methods. In the most important case of weakly transparent barrier (tunnel regime), we have restricted our consideration by the model approach in which only zero harmonic (i.e., the average time value) of the function $\check G$ is taken into account, since its higher harmonics with numbers $m=1,2,\dots$ decrease as $D^m$ \cite{BezuglyRC06, BezuglySUST}. In fact, such approach can be considered as an attempt to describe, at least qualitatively, the coherent MAR in clearly tractable terms of the local density of states and the distribution function, usually applied to the incoherent MAR regime \cite{Bez2000}. In this approximation, the dc quasiparticle current is expressed through the following integral over the energy $E$,
\begin{eqnarray}
I =  \int_0^{eV}\frac{\rmd E}{e R}  J(E), \quad J = \sum\nolimits_{k=-\infty}^{\infty} j_k, \label{I002}
\\
j_k = \left( n_{k-1} - n_k \right) \rho_k^{-1}, \quad \rho_k^{-1} = N_k N_{k-1}. \label{jrho}
\end{eqnarray}
Here $R$ is the junction resistance, $N_k =N(E_k)$ is the quasiparticle density of states in the junction area, normalized to its value in the normal metal, $E_k=E+keV$, $n_k =n(E_k)$ is the non-equilibrium distribution function of quasiparticles satisfying the following recurrence relation
\begin{equation}
\Theta(|E_k| - \Delta) \left[n_k - n_{\rm F}(E_k)\right] = r (j_k -
j_{k+1}), \quad r = R_N/R.\label{recurr}
\end{equation}
Here $n_F(E)$ is the equilibrium Fermi function, $R_N$ is the resistance of the junction leads in the normal state, and $\Theta(x)$ is the Heaviside step function.

\begin{figure}[tb]
\begin{center}
\epsfxsize=8cm\epsffile{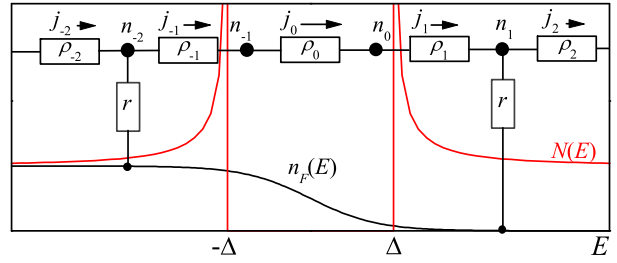}
\end{center}
\caption{Equivalent circuit representation of the MAR transport in particular case of the 3-particle current.}
\label{circuit}
\end{figure}

Equations \eref{I002}-\eref{recurr} have a clear interpretation in terms of an equivalent infinite electrical circuit in the energy space with the period $eV$ (Figure \ref{circuit}). According to \eref{I002}, the current spectral density $J(E)$ is the sum of the partial currents $j_k$ flowing through the chain of the tunnel ``resistors'' $\rho_k$ which connect adjacent chain nodes with the effective ``potentials'' $n_k$. At $|E|>\Delta$, the nodes are connected to the distributed ``voltage source'' $n_F(E)$ through the lead ``resistors'' $r_k$; from this viewpoint, the recurrence relation \eref{recurr} has the meaning of the Kirchhoff rules for the partial currents. We note that due to full Andreev reflection, the nodes inside the gap ($|E_k|<\Delta$) are disconnected from the equilibrium voltage source (bulk electrodes), therefore the subgap quasiparticle population is highly nonequilibrium, and all the currents through the subgap tunnel resistors are equal.

Since $R_N$ is usually smaller than $R$, we have neglected in \cite{BezuglyRC06,BezuglySUST} the resistors $r$ which implies equilibrium population in the junction leads. In this approximation, the current $I(V)$ and the subgap distribution function $n(E)$ at arbitrary temperatures read
\begin{eqnarray}\label{Inet}
I(V) &=  \int_0^{eV}\frac{\rmd E}{eR}(M_- + M_+)j_\Delta
\\
\nonumber &+ 2\int_\Delta^\infty\frac{\rmd E}{eR}{[n_{\rm F}(E) - n_{\rm
F}(E_1)]}{N(E)N(E_1)},
\end{eqnarray}
\begin{equation} \label{nnet}
n(E)=n_+ + (n_- - n_+)\rho_+/\rho_\Delta.
\end{equation}
where  $j_\Delta(E) = (n_- - n_+)/\rho_\Delta$ is the spectral density of the current flowing through the chain of resistors crossing the gap (shortly - ``subgap resistors''), $n_\pm = n_{\rm F}(E\pm  eV M_\pm)$ are the Fermi functions, the integers $\pm M_\pm$ (where $M_\pm (E) = 1 + {\rm Int}\,\left[(\Delta \mp E)/eV\right]$) are the indexes of the nodes outside the gap nearest to its edges, $\mathrm{Int}(x)$ is the integer part of $x$, and
\begin{eqnarray}\label{rhoDelta}
\rho_\Delta = \sum\nolimits_{k= 1-M_- }^{M_+} \rho_k, \quad  \rho_+ = \sum\nolimits_{k=1}^{M_+}\rho_k.
\end{eqnarray}
are the net subgap resistance and the resistance of the subgap MAR chain at the right side of the resistor $\rho_0 \equiv \rho(E)$, respectively. The second term in \eref{Inet} is the current of thermally excited quasiparticles, and the first term is the current flowing through the subgap resistors; the number of them, $M_+ + M_-$, gives the value of electric charge (in units of $e$) transferred during one multiparticle tunneling event. Thus, the chain with only one resistor crossing the gap (which is possible only at $eV >2\Delta$) describes the single-particle tunneling, and the first term in \eref{Inet} is reduced to the standard result of the tunnel model \cite{Werthamer},
\begin{eqnarray}\label{IBCS}
I =  \int_\Delta^{eV-\Delta}\frac{\rmd E}{eR}{[n_{\rm F}(E_{-1}) - n_{\rm
F}(E)]}{N(E)N(E_{-1})}.
\end{eqnarray}

As the voltage decreases, the number of the subgap resistors increases, which manifests emergence of multiparticle processes; for example, Figure \ref{circuit} illustrates the 3-particle current. In this case, some nodes get into the subgap region where the BCS density of states is zero, which leads to divergence of the subgap tunnel resistance $\rho_\Delta$ and, correspondingly, to disappearance of the subgap current. Thus, in order to calculate the latter, one has to find tunnel corrections to $N(E) = \re \cosh\theta$ using the recurrence relation for the spectral angle $\theta(E)$ derived in \cite{BezuglySUST},
\begin{eqnarray}
&\rmi \sinh[\theta(E) - \theta_{\rm s}(E)] = {\W} \sinh\theta_{\rm s}(E) \sinh\theta  \nonumber
\\
&\times[\cosh\theta(E+eV) + \cosh\theta(E-eV)],  \label{Eqtheta}
\end{eqnarray}
where $\theta_{\rm s}(E) = {\rm arctanh}(\Delta/E)$ is its unperturbed (BCS) value. This results in a ladder-like structure of $N(E)$ (see Fig.~4 in \cite{BezuglyRC06}), which penetrates into the energy gap from its edges by steps of the lengths $eV$ and the heights scaled by the transparency parameter
\begin{equation}\label{Wtilde}
\W = ({3\xi_0^2}/{4\ell d}) D.
\end{equation}
In a diffusive planar junction with thin junction plates, $d \sim \ell \ll \xi_0$, the value of $\W$ may greatly exceed the bare transmission coefficient $D$, and the SGS scaling was found to be similar to the one in the ballistic junction with $D_\mathrm{eff}=4\W$. Similar enhancement of the effective transmission coefficient appears in the 1D geometry (the tunnel contact between the edges of the leads) which has been found earlier for the dc Josephson current in superconducting tunnel junctions \cite{BBG} and for the subgap current in a normal metal (semiconductor)/insulator/superconductor voltage biased junction \cite{Wees1992}. In this case, the enhancement effect is smaller, $\sim \xi_0/\ell$; nevertheless, all results of our paper are also applicable.

\section{Nonequilibrium in junction leads}\label{SecNoneq}

The nonequilibrium in the superconducting leads is produced by the tunnel injection  of excess quasiparticles. This effect is generally rather small since the diffusion of nonequilibrium quasiparticles away from the junction is rapid compared to the tunneling rate, that is reflected in smallness of the diffusion resistance $r \ll 1$. However, the energy-dependent tunnel resistances $\rho_k$ may become anomalously small at some singular points of the product $N_k N_{k-1}$ (which, according to the Fermi golden rule, enhances the tunneling probability) and thus may be comparable with $r$. At small $r$ and zero temperatures, the tunnel currents outside the gap rapidly decrease as the distance from the gap edges grows, therefore it is enough to keep only one or two side resistors $r$ near the gap edge.

First we address the role of the nonequilibrium in formation of the excess current $I_{\mathrm{exc}}$, i.e., voltage-independent deviation of the total current from the ohmic CVC at large voltage, $eV \gg \Delta$. In our previous paper, only the contribution $I_2$ of the two-particle processes to $I_{\mathrm{exc}}$ at $r=0$ has been evaluated (Eq.(58) in \cite{BezuglySUST}); however, the net excess current also involves contribution from the single-particle current $I_1$ described by \eref{IBCS}. Taking the functions $N(E)$ in this equation within the next approximation in $\W$ by using the improved perturbation theory \cite{BezuglySUST} for the solution of \eref{Eqtheta} near the singularities of $N(E)$, we found this contribution to be negative (a deficit current) and twice larger than the contribution of the two-particle current. As the result, the net CVC demonstrates the deficit current
\begin{equation}\label{Iexcr=0}
I_{\mathrm{exc}} = -\frac{\Delta}{eR}\sqrt{2\W}, \quad r=0.
\end{equation}
\begin{figure}[b]
\begin{center}
\epsfxsize=8cm\epsffile{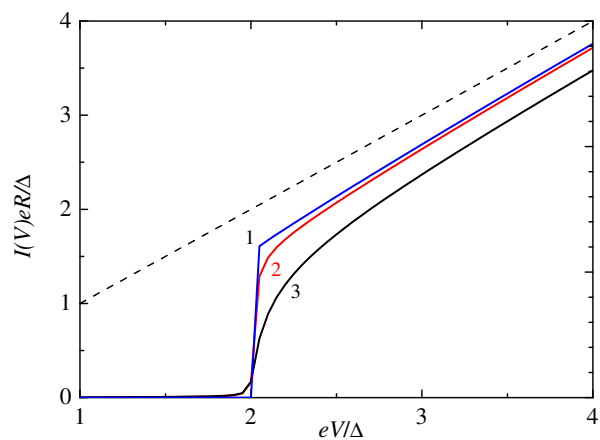}
\end{center}
\caption{CVC near the onset of the single-particle current at $\W=0.001$, $T=0$: the tunnel model result (curve 1); our result for $r=0$ (curve 2) and $r = 0.05$ (curve 3); Ohm's law (dashed line).}
\label{onset}
\end{figure}

At finite $r$, this estimate may change because under certain conditions, this parameter can play the role of the cut-off factor in the tunnel corrections to $I_1$. Keeping one side resistor $r$ at both sides of the gap, we obtain
\begin{eqnarray}\label{I_r}
I_1 = \frac{1}{eR}\int_\Delta^{eV-\Delta} \frac{dE}{2r +(N_0
N_{-1})^{-1}},
\\
I_2 = 4 \int_0^{\Delta} \frac{dE}{eR} \frac{1}{2r + (N_0
N_{-1})^{-1} + (N_0 N_{1})^{-1}}. \label{I_r2}
\end{eqnarray}
In Eq.\eref{I_r}, we can take $N_0$ and $N_1$ in the BCS form, $N_{\rm s}(E) = \re \cosh \theta_{\rm s}= |E|\Theta(|E|-\Delta)/\sqrt{E^2-\Delta^2}$, then the contribution $I_1$ to the excess current is easily evaluated at $r \ll 1$,
\begin{equation}\label{IexcI1}
I_{\mathrm{exc}}^{(1)} = -\frac{4\Delta}{eR}r\left(\ln\frac{1}{r}-1\right).
\end{equation}
Evaluation of $I_2$ is more complicated because $N_0$ in Eq.\eref{I_r2} is to be calculated inside the energy gap by solving equation \eref{Eqtheta} within the linear approximation in $\W$,
\begin{equation}\label{DOSSPT}
N(E)=\W({1 - E^2/\Delta^2})^{-3/2} [N_{\rm s}(E+eV)+N_{\rm s}(E-eV)],
\end{equation}
or by the nonperturbative expansion of $N(E)$ in the vicinity of the singular point $E=\Delta$ \cite{BezuglyRC06},
\begin{equation}\label{DOSIPT}
N(E) = \frac{1}{2\sqrt{\W}} \im \frac{1}{\sqrt{\epsilon -i}}, \quad \epsilon = \frac {\Delta - E}{2\W\Delta}.
\end{equation}
Our analysis shows that the simple approximation \eref{DOSSPT} is applicable when $\W \ll r^2$ which yields the value
\begin{equation}\label{IexcSPT}
I^{(2)}_{\mathrm{exc}} = 3.84 \Bigl(\frac{\W^2}{r}\Bigr)^{1/3}\frac{\Delta}{eR}
\end{equation}
smaller than the negative contribution \eref{IexcI1} of the single-particle current. In the opposite case, $\W \gg r^2$, the parameter $r$ can be neglected both in $I_1$ and $I_2$, and we return to the value of the excess current \eref{Iexcr=0} calculated at $r=0$. From this we conclude that in our model the excess current is always negative.

Numerical calculation shows that at small enough transparency parameter $\W$, the nonequilibrium in the junction leads noticeably changes not only the excess current but also the overall CVC shape at $eV>2\Delta$ as compared with the usual tunnel model formula \eref{IBCS} taken with the BCS density of states, $N=N_{\rm s}$. As shown in Figure \ref{onset}, the jump of the single-particle current at its threshold $eV=2\Delta$ essentially smoothes and acquires a finite slope. We note that the values of the excess current calculated above are reached within the appropriate accuracy only at very large voltages, $eV \gtrsim 10^2 \Delta$; in particular, the ``deficit current'' in curve 1 (tunnel model result) is actually fictitious and disappears at large enough voltages.

At small enough $\W$, the nonequilibrium in the junction leads also affects multiparticle currents at $eV \leq 2\Delta$, especially in the vicinity of their peaks. For instance, at $eV=2\Delta$, where the 2-particle current reaches a maximum value, one must modify the expression \eref{I_r2}, where the partial current $j_{-1}$ flowing through the resistor $r$ and anomalously small resistor $\rho_{-1}$ must be taken into account. This leads to the following relation
\begin{equation}\label{I2Dn}
I_2 = \frac{5}{eR}\int_0^\Delta \frac{dE}{3r/2 + N_0 N_{-1} + N_0 N_1}.
\end{equation}
Using the approximation \eref{DOSSPT} for the subgap value of $N_0(E)$, we finally obtain the result relevant for $r \gg \W^{4/3}$,
\begin{equation}\label{I2peak}
I_2(2\Delta) \approx \frac{4\Delta}{eR}\left( {\frac{\W^4}{r^3}}\right) ^{1/7}.
\end{equation}

The case of the 3-particle current is conceptually similar to the previous one: we have to consider two circuit segments, the subgap segment consisting of the three resistors, $\rho_{-1}$, $\rho_0$ and $\rho_1$, and an additional one containing $\rho_{-2}$ since this resistance becomes anomalously small at $eV = \Delta$. This results in the following equation,
\begin{equation}\label{I3Dn-eq1}
I_3 = \frac{7}{eR}\int_{eV/2}^\Delta \frac{dE}{3r/2 + \rho_{-1} +
\rho_0 + \rho_1}.
\end{equation}
Within the approximation \eref{DOSSPT} for $N$ and $N_{-1}$, we obtain the peak value of $I_2$ at $eV=\Delta$,
\begin{equation} \label{I3peak}
I_3(\Delta)  \approx 5.9\frac{\Delta}{eR}\left(
{\frac{\W^8}{r^3}}\right) ^{1/7}.
\end{equation}
Comparing equations \eref{I2peak} and \eref{I3peak} with the results of \cite{BezuglySUST}, we see that the nonequilibrium in the junction leads noticeably suppresses the peaks of the multiparticle currents.

\section{Nonequilibrium and inelastic relaxation inside the gap}\label{SecRelax}

As noted above, the quasiparticle distribution function $n(E)$ in the subgap region, $|E|<\Delta$, is far from equilibrium, because the subgap quasiparticles are disconnected from the equilibrium reservoirs. A consistent analysis of such nonequilibrium state requires consideration of inelastic relaxation processes \cite{Arutyunov}. We will model the inelastic scattering by adding the collision term in the $\tau$-approximation to the diffusive kinetic equation introduced in \cite{BezuglySUST},
\begin{equation}\label{kin}
\nabla(D_+ \nabla n) = N \frac{n-n_F}{\ell_\epsilon^2},
\end{equation}
where $\ell_\epsilon = \sqrt{\mathcal{D}\tau_\epsilon}$ and $\tau_\epsilon$ are the inelastic relaxation length and time, respectively; $D_+$ is the energy-dependent dimensionless diffusion coefficient \cite{BezuglySUST}. Thus, in presence of this term, the spectral current $j(E) = -D_+ \nabla n$ is not conserved inside the gap.

It is possible to include the relaxation effect into the circuit scheme by the method suggested by Volkov \cite{VolkovSIN} and used in \cite{BezuglySUST} for derivation of the recurrence relation \eref{recurr}. Assuming $n(\bm{r})\approx \const$ within the junction plates, integrating \eref{kin} over the volume of the bottom plate (a similar procedure applies to the top plate) and taking into account that at the distance $\gtrsim\xi_0$ from the junction, all spectral characteristics approach their BCS values (in particular, $D_+$ turns to zero in the subgap region), we obtain the boundary value of the spectral current at the bottom side of the barrier,
\begin{equation}\label{kin1}
D_+ \partial_y n\bigr|_{-0} = d N \frac{n-n_F}{\ell_\epsilon^2}
\end{equation}
where the $y$ axis is perpendicular to the contact plane. Substituting \eref{kin1} to the boundary condition for the distribution function (equation (20) in \cite{BezuglySUST}), we obtain the recurrence relation for $n(E)$ at $|E_k|<\Delta$,
\begin{eqnarray}\label{bound}
n_F(E_k)-n_k = r_{\epsilon k} (j_{k+1}-j_{k}), \\
\nonumber r_\epsilon = 4\W \tau_\epsilon
\Delta {N}^{-1} = A\W {N}^{-1},\quad  A = 4\tau_\epsilon \Delta,
\end{eqnarray}
where the parameter $A$ is usually large, $A \gg 1$, for a standard BCS superconductor. In the circuit terms, this equation describes leakage of nonequilibrium quasiparticles from the subgap nodes to the equilibrium source through the resistors $r_\epsilon$. This modifies the equation \eref{Inet} for the electric current, as well as the expression \eref{nnet} for the distribution function, because the partial currents may flow not only through the tunnel resistors $\rho_k$ but also through the leakage resistors $r_{\epsilon k}$; therefore, as noted above, the subgap partial currents $j_k$ are not equal. The magnitude of this effect depends on the ratio between $r_\epsilon$ and adjacent tunnel resistors, i.e. between the tunneling and relaxation rates. In what follows, we restrict ourselves for simplicity by the case $T=0$, when $n_F(E)$ is a step-like function, and neglect the effect of nonequilibrium outside the gap described in the previous section.

We start our consideration from the analysis of the two-particle current (obviously, the single-particle current is not affected by the subgap relaxation). Solving the corresponding circuit with the leakage resistor attached to the subgap node $k=0$, we obtain the relevant partial currents and the net spectral current $j^{(2)}(E)$,
\begin{eqnarray}\label{I2E}
j_{0}=g_{0}\go_1/G, \quad j_1 =
g_{0}g_1/G, \\ \nonumber G = g_0 +\go_{1},\quad \go_1 = g_1+g_{\epsilon 0} \\
j^{(2)}(E) = j_0 +j_1 = \frac{1+g_1 \ro_1}{\rho_0 + \ro_1}, \quad \ro_k =
\frac{1}{\go_k},
\end{eqnarray}
where $g_k = \rho_k^{-1}$, $g_{\epsilon k} = r_{\epsilon k}^{-1}$ are the conductances of the network resistors. As follows from \eref{bound}, the characteristic magnitude of $g_{\epsilon 0}$ is determined by the parameter $A^{-1}$ (we remind that at $eV > \Delta$, ${N(E)} \sim \W$ in the subgap region). Thus, the ``inelastic leakage'' can be neglected, if $g_{\epsilon 0} \ll g_{0,1} \sim \W$, i.e., at $A\W \gg 1$ (weak relaxation). In the opposite case, $A\W \ll 1$ (strong relaxation), the partial current $j_1$ is ``short-circuited'' by the comparatively small leakage resistor $r_{\epsilon 0}$, and therefore $j_1$ can be neglected as compared to $j_{0} \approx g_{0}$. As the result, the electric current spectral density in the strong relaxation limit, $j_0 + j_{1} \approx 1/\rho_{0}$, insignificantly differs from its value $j_0 + j_{1} \approx 2/(\rho_{1}+\rho_0)$ in the collisionless limit because the resistances $\rho_0$ and $\rho_1$ are of the same order.

\begin{figure}[th]
\begin{center}
\epsfxsize=7.5cm\epsffile{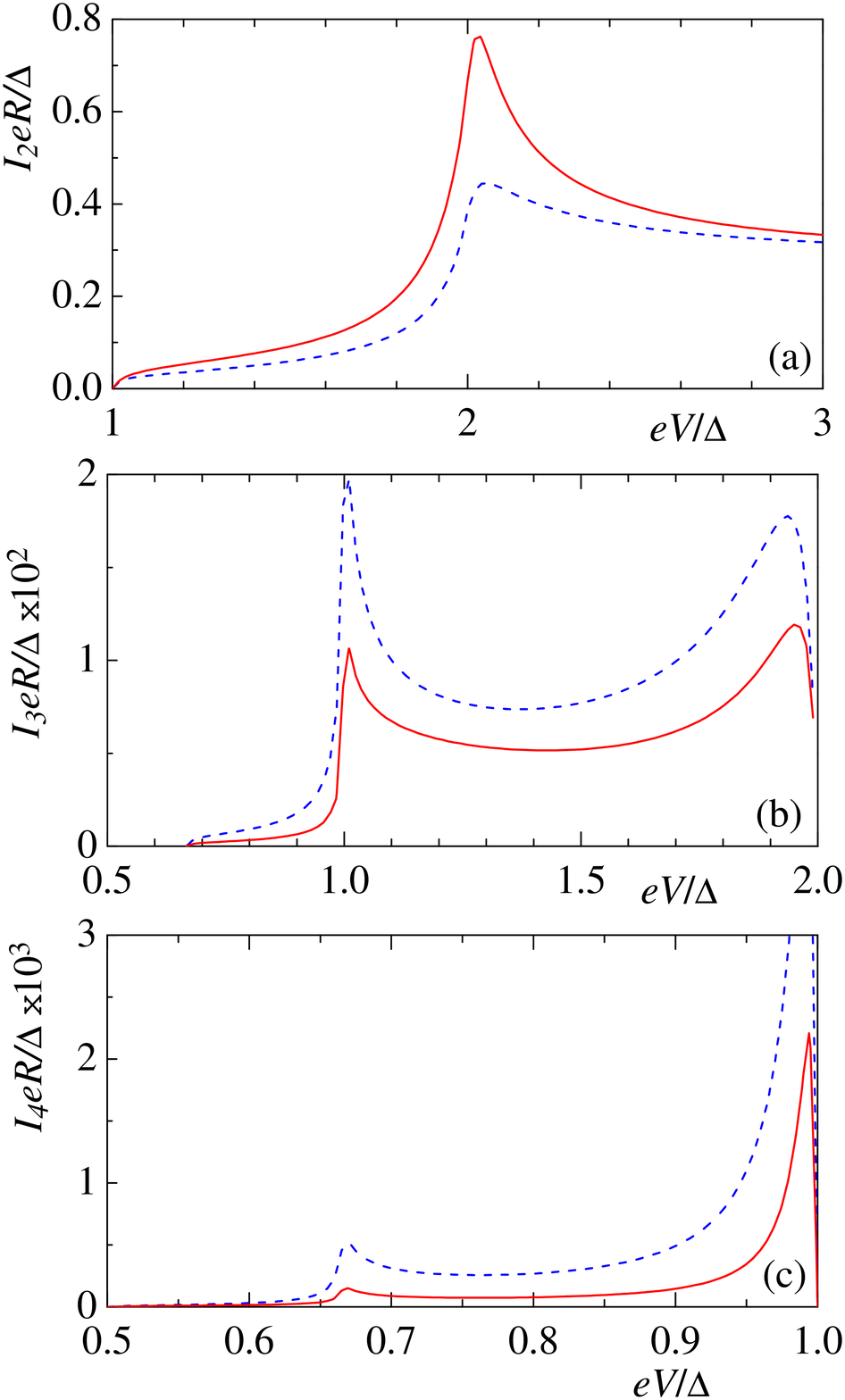}
\end{center}
\caption{Voltage dependences of the 2-, 3- and 4-particle currents (panels (a), (b) and (c), respectively) at $\W=0.01$ in the absence of relaxation (dashed lines) and in the strong relaxation limit (solid lines).} \label{relax}
\end{figure}

The effect of inelastic collisions is most essential for higher $n$-particle currents, $n > 2$. For odd $n$, the current through the central resistor $\rho_0$ dominates; other tunnel resistors are shortened by the leakage resistors surrounding $\rho_0$. Thus, the estimate of the spectral current is $1/\rho_0$, while in the collisionless limit, it has the value $n/\rho_\Delta$ (where $\rho_\Delta \approx \rho_0$), i.e. $n$ times larger. For even $n$, one of two largest central resistors is also shortened, which leads to a similar estimate, $1/\rho_0$, for the spectral current. However, in the collisionless limit, the estimate for $\rho_\Delta \approx \rho_0+\rho_1$ is $2\rho_0$, i.e. the spectral current is $n/2$ times larger (note that this estimate is also formally applicable to the 2-particle current).

A more detailed analysis shows that under the condition $A\W^2 \ll 1$, which is always satisfied for realistic values of the parameter $A \sim 10^2-10^3$ and $\W \lesssim 10^{-2}$, the inelastic suppression of the 3- and 4-particle currents described above develops only in the strong relaxation regime, $A\W \ll 1$, while in the opposite limit, $A\W \gg 1$, the relaxation weakly affects the CVC at $eV > \Delta/2$. This is not the case for higher currents; for $n>4$, suppression of the multiparticle currents described above begins in the weak relaxation regime.

In order to verify these qualitative considerations, we performed numerical calculations of several multiparticle currents in the strong and weak relaxation regimes, using \eref{I2E} and similar full analytical expressions for the spectral densities of 3- and 4-particle currents,
\begin{eqnarray}\label{I4E1}
j^{(3)}(E) = \frac{1+g_1 \ro_1 + g_{-1}\ro_{-1}}{\rho_0
+\ro_1+\ro_{-1}},
\\
j^{(4)}(E)
\\ \nonumber
= \frac{2+g_{-1}\ro_{-1} + g_2 \ro_2
+g_{E0}(\rho_1+\ro_2)(1+g_{-1}\ro_{-1})}{\rho_0+\ro_{-1}+\rho_1 +
\ro_2 +g_{E0}(\rho_0 + \ro_{-1})(\rho_1 +\ro_2)}.
\end{eqnarray}
The results shown in Figure \ref{relax} qualitatively confirm our preliminary estimates: in the case of strong relaxation, the two-particle current changes weakly and even slightly grows; the 3-particle current noticeably decreases (but smaller than predicted above); the 4-particle current is indeed suppressed by 2 times and even more. The deviations from the qualitative estimates can be explained by the nontrivial energy dependence of the tunnel resistances and enhanced contributions of the vicinities of the singular points, where the simple estimates of $\rho_k$, based on the perturbative formula \eref{DOSSPT}, may appear to be too rough.

Similar considerations can be applied to the analysis of the distribution function $n(E)$. In the absence of the inelastic scattering, equation \eref{nnet} determines $n(E)$ as the ``potential'' of the node with the index $k=0$ of the ``voltage divider'', consisting of all subgap resistors and connected to the sources with the voltage difference $n_- - n_+$. For odd-particle currents, one central subgap resistor greatly exceeds the resistance of other elements, therefore the ``potentials'' at its left (right) edges, as well as at other nodes at the left (right) side of this resistor, are approximately equal to $n_-$ ($n_+$), respectively. From this we conclude that within the energy/voltage interval where the odd-particle current exist, the distribution function is close to the value of the equilibrium Fermi function in the vicinity of the nearest edge of the energy gap. Similar conclusion can be made for even-particle currents, where two central resistors dominate, except the case when the node $k=0$ appears between these two resistors; obviously, in such situation, the value of $n(E)$ can be estimated as an average between $n_\pm$, i.e. close to $1/2$. As the voltage decreases, the size of the dominating resistors in the energy space, i.e. the size of the nonequilibrium energy range, $|E| \lesssim eV$, gradually shrinks and finally disappears inside the temperature smearing of the quasiparticle distribution; apparently, this process is accelerated by the inelastic relaxation.

\begin{figure}[t]
\begin{center}
\epsfxsize=8cm\epsffile{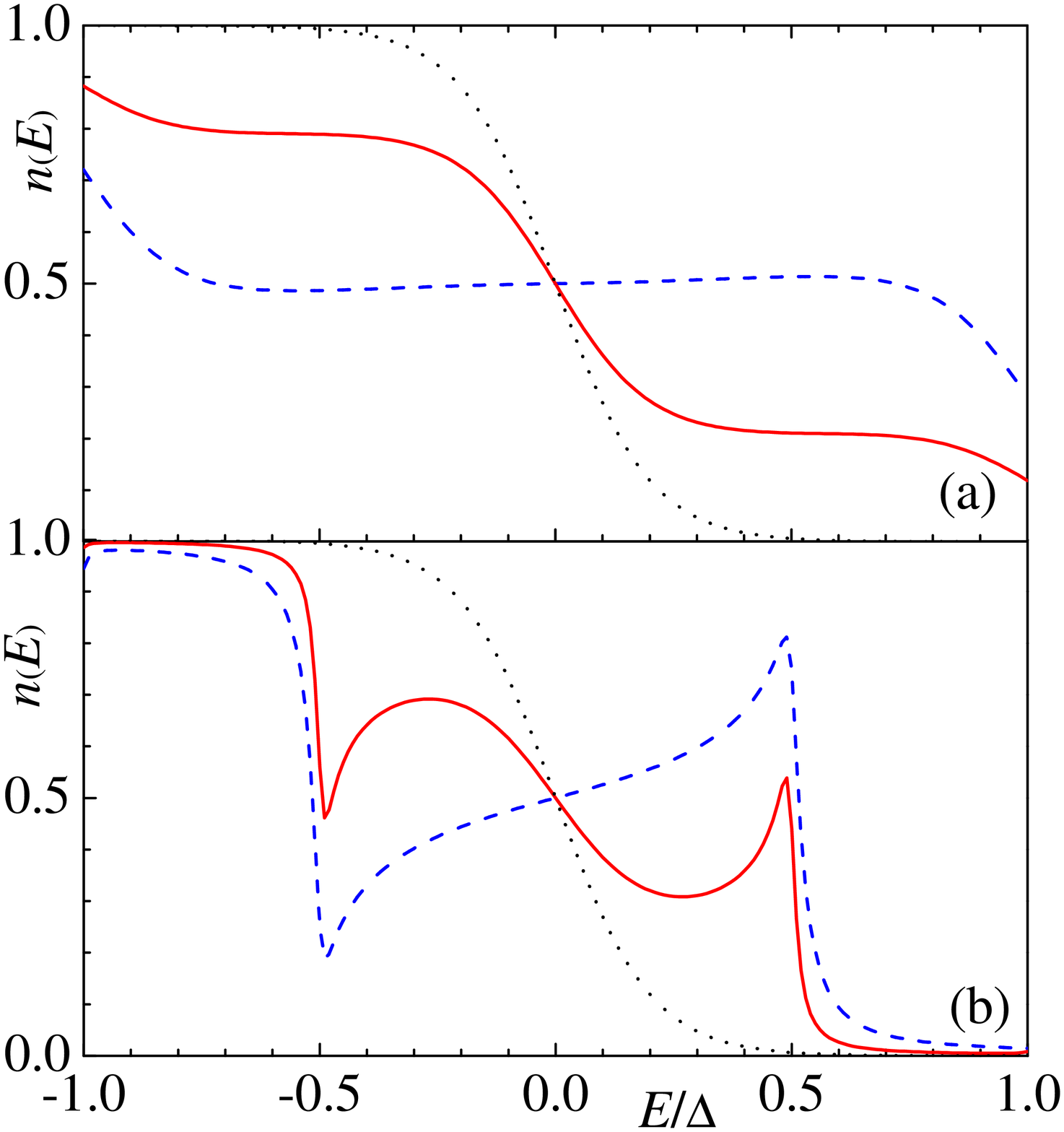}
\end{center}
\caption{Subgap quasiparticle distributions at $T=0.1\Delta$, $\W=0.01$ and different applied voltages: $eV = 2.5\Delta$ (a) and $1.5\Delta$ (b) in the weak ($A\W = 5$, dashed lines) and strong ($A\W = 0.3$, solid lines) inelastic relaxation regimes; the equilibrium distribution is shown by dotted lines.} \label{n_relax}
\end{figure}

Using the solutions for the partial currents $j_k$  and the relation \eref{bound}, we obtained exact expressions for $n(E)$ in presence of the inelastic relaxation and calculated the shape of the distribution function. The results for different applied voltages are shown in Figure \ref{n_relax}; for more clarity, we plotted $n(E)$ at low but finite temperature. Panel (a) demonstrates $n(E)$ at the voltages above the threshold of the single-particle current, $eV > 2\Delta$, when the subgap population is formed by the 2-particle processes within the whole subgap region, $|E|< \Delta$. In accordance with our qualitative analysis, $n(E)$ is almost constant in the weak relaxation limit and close to $0.5$ (dashed line); indeed, both subgap conductances, $N_0 N_1$ and $N_0 N_{-1}$, differ only by the factors $N_{\pm 1}$ which are the BCS densities of states above the gap. At large enough applied voltage chosen in Figure \ref{n_relax}(a), the difference between them is rather small and leads to minor deviations of $n(E)$ from the constant value. Interestingly, these deviations correspond to a partially inverted population of the subgap region. As $eV$ approaches $2\Delta$, the inversion effect enhances because the difference between the values of $N_{\pm 1}$ grows, especially at the edges of the subgap regions, where one of them has a singular point. Such effects are more pronounced in the voltage region $\Delta < eV < 2\Delta$, when the subgap population is created by the 2-particle current within the restricted area $|E| < eV -\Delta$, as shown in Figure \ref{n_relax}(b). At the edges of this energy interval, one of the  subgap resistors becomes anomalously small (namely, the left resistor at $E = eV -\Delta$ and the right one - at $E = \Delta-eV$), which leads to the enhancement of the inversion effect. At larger $|E|$, the population is determined by the 3-particle current and becomes close to the equilibrium distribution, in accordance with our estimates. The relaxation role is obvious: it brings the distribution function closer to the equilibrium one, which is illustrated by solid curves in Figure \ref{n_relax}. We would like to mention that the ``subgap nonequilibrium'' studied in this section is closely related to the ``injection nonequilibrium'' in the double-barrier junctions studied in \cite{Brinkman2003, Brinkman2000, Kupriyanov1999, Heslinga1993}.

\section{Summary}\label{SecSummary}

In conclusion, we have analyzed theoretically the influence of the relaxation processes on the nonequilibrium quasiparticle distributions in voltage biased diffusive tunnel junctions fairly distant from massive equilibrium electrodes. In this case the mesoscopic approach (see, e.g., \cite{Nazarov}), based on averaging of the result for a single ballistic quantum channel over the distribution of junction transparencies in the normal state, is inapplicable. Using the circuit theory approach introduced in our previous papers \cite{BezuglyRC06, BezuglySUST}, we have included the relaxation factors in our circuit scheme as additional resistive elements whose resistances reflect the characteristic times of the inelastic or diffusive relaxation.

As far as the diffusive escape from the junction plates to its leads is relevant only for nonequilibrium quasiparticles with energies outside the gap (the subgap quasiparticles are locked inside the plates), the most prominent effect of the diffusive relaxation appears at $eV > 2\Delta$. Namely, while the relaxation intensity decreases, the CVC exhibits crossover from the result of a simple tunnel model, with a sharp jump of the single-particle current at the threshold $eV = 2\Delta$, to a smooth voltage dependence with the finite slope and comparatively large deficit current. The multiparticle currents, as well as the subgap distribution function, are most sensitive to inelastic relaxation of the subgap quasiparticles. We discuss the shape of the subgap distribution function which may be inversive within a certain energy/voltage range and found that strong inelastic relaxation significantly reduces the magnitude of the $n$-particle currents, approximately by $n$ times for odd-, and by $n/2$ for even-particle ones.

\ack{The article was prepared within the framework of the Academic Fund Program of the National Research University Higher School of Economics (HSE) in 2016  (grant No. 16-05-0029 ``Physics of low-dimensional quantum systems'') and supported within the framework of a subsidy granted to the HSE by the Government of the Russian Federation for the implementation of the Global competitiveness program.}


\section*{References}

\begin{thebibliography}{99}

\bibitem{BezuglyRC06}
Bezuglyi E V, Vasenko A S, Bratus' E N, Shumeiko V S and Wendin G 2006 \PR B
\textbf{73} 220506(R)

\bibitem{BezuglySUST}
Bezuglyi E V, Vasenko A S, Bratus' E N, Shumeiko V S and Wendin G 2007 \SUST \textbf{20} 529

\bibitem{MPT}
Schrieffer J R and Wilkins J W 1963 \PRL {\bf 10} 17;
Wilkins J W 1963 {\em Tunnelling Phenomena in Solids} (New York: Plenum) p 333;
Hasselberg L E, Levinsen M T and Samuelsen M R 1974 \PR B {\bf 9} 3757

\bibitem{KBT}
Klapwijk T M, Blonder G E and Tinkham M 1982 {\it Physica} B+C {\bf 109-110}
1657

\bibitem{Bratus95}
Bratus' E N, Shumeiko V S and Wendin G 1995 \PRL {\bf 74} 2110

\bibitem{Averin95}
Averin D and Bardas A 1995 \PRL {\bf 75} 1831

\bibitem{Cuevas96}
Cuevas J C, Martin-Rodero A and Yeyati A L 1996 \PR B {\bf 54} 7366

\bibitem{Jan2000}
Ludoph B, van der Post N, Bratus' E N, Bezuglyi E V, Shumeiko V S, Wendin G
and van Ruitenbeek J M 2000 \PR B {\bf 61} 8561; Naaman O and Dynes R C 2004
\SSC {\bf 129} 299

\bibitem{Gubrud2001}
Patel V and Lukens J E 1999 {\it IEEE Trans. Appl. Supercond.} {\bf 9} 3247;
Gubrud M A, Ejtnaes M, Bercley A J, Ramos R C (Jr), Anderson J R, Dragt A J,
Lobb C J and Wellstood F C 2001 {\it IEEE Trans. Appl. Supercond.} {\bf 11}
1002; Lang K M, Nam S, Aumentado J, Urbina C and Martinis J M 2003 {\it IEEE
Trans. Appl. Supercond.} {\bf 13} 989; Oh S, Cicak K, McDermott R, Cooper K B,
Osborn K D, Simmonds R W, Steffen M, Martinis J M and Pappas D P 2005 \SUST
{\bf 18} 1396

\bibitem{Makhlin}
Makhlin Yu, Schon G, and Shnirman A 2001 \RMP \textbf{73} 357

\bibitem{Wendin}
Wendin G and Shumeiko V 2007 {\it Low Temp. Phys.} \textbf{33} 724

\bibitem{Paauw}
Paauw F G, Fedorov A, Harmans C J P M and Mooij J E 2009 \PRL \textbf{102} 090501

\bibitem{Martinis}
Martinis J M, Ansmann M and Aumentado J 2009 \PRL \textbf{103} 097002;
Wenner J, Yin Y, Lucero E, Barends R, Chen Y, Chiaro B, Kelly J, Lenander M, Mariantoni M, Megrant A, Neill C, O'Malley P J J, Sank D, Vainsencher A, Wang H, White T C, Cleland A N and Martinis J M 2013 \PRL \textbf{110} 150502

\bibitem{Sandberg}
Sandberg M, Wilson C M, Persson F, Bauch T, Johansson G, Shumeiko V, Duty T and Delsing P 2009 {\it Appl. Phys. Lett.} \textbf{92} 203501

\bibitem{Harvey}
Harvey T J, Rodrigues D A and Armour A D 2008 \PR B \textbf{78} 024513

\bibitem{Levenson-Falk}
Levenson-Falk E M, Kos F, Vijay R, Glazman L and Siddiqi I 2014 \PRL \textbf{112} 047002

\bibitem{Pekola1}
Pekola J P, Vartiainen J J, Mottonen M, Saira O-P, Meschke M and Averin D V
2007 {\it Nature Phys.} \textbf{4} 120

\bibitem{Muhonen}
Muhonen J T, Meschke M and Pekola J P 2012 {\it Rep. Prog. Phys.} \textbf{75} 046501

\bibitem{Rajauria}
Rajauria S, Courtois H and Pannetier B 2009 \PR B \textbf{80} 214521

\bibitem{VH}
Vasenko A S and Hekking F W J  2009 {\it J. Low Temp. Phys.} \textbf{154} 221

\bibitem{VBCH}
Vasenko A S, Bezuglyi E V, Courtois H and Hekking F W J 2010 \PR B \textbf{81} 094513

\bibitem{Kawabata}
Kawabata S, Ozaeta A, Vasenko A S, Hekking F W J and Bergeret F S 2013 {\it Appl. Phys. Lett.} \textbf{103} 032602

\bibitem{Marcus}
Marcus S M 1966 \PL {\bf 19} 623; Marcus S M 1966 \PL \textbf{20} 236; Rowell J M and Feldmann W L 1968 \PR B \textbf{172} 393

\bibitem{Kle}
Kleinsasser A W, Miller R E, Mallison W H and Arnold G B 1994 \PRL \textbf{72}
1738

\bibitem{Naveh}
Naveh Y, Patel V, Averin D V, Likharev K K and Lukens J E 2000 \PRL \textbf{85} 5404

\bibitem{Greibe}
Greibe T, Stenberg M P V, Wilson C M, Bauch T, Shumeiko V S and Delsing P 2011
\PRL \textbf{106} 097001

\bibitem{VolkovSIN}
Volkov A F 1994 {\it Physica} B {\bf 203} 267

\bibitem{Berthod}
Berthod C and Giamarchi T 2011 \PR B \textbf{84}, 155414

\bibitem{LOnoneq} Larkin A I and Ovchinnikov Yu N 1986 {\it
Nonequilibrium Superconductivity} ({\it Modern Problems in Condensed Matter
Sciences} vol 12) ed D N Langenberg and A I Larkin (Amsterdam: North-Holland)
p 493

\bibitem{Belzig}
Belzig W, Wilhelm F K, Bruder C, Sch\"{o}n G and Zaikin A D 1999 {\it
Superlatt. Microstruct.} {\bf 25} 1251; Eschrig M 2000 \PR B {\bf 61} 9061

\bibitem{Bezugly2005}
Bezuglyi E V, Vasenko A S, Shumeiko V S and Wendin G 2005 \PR B \textbf{72} 014501

\bibitem{Bez2000} Bezuglyi E V, Bratus' E N, Shumeiko V S, Wendin G and Takayanagi H 2000 \PR B \textbf{62} 14439

\bibitem{Werthamer}
Werthamer N R 1966 \PR \textbf{147} 255; Larkin A I and Ovchinnikov Yu N 1967
{\it Sov. Phys.--JETP} \textbf{24} 1035

\bibitem{BBG}
Kupriyanov M Yu 1992 {\it JETP Lett.} {\bf 56} 399; Bezuglyi E V, Bratus' E N
and Galaiko V P 1999 {\it Low Temp. Phys.} {\bf 25} 167; Galaktionov A V and
Chang-Mo Ryu 2000 \JPCM {\bf 12} 1351

\bibitem{Wees1992}
van Wees B J, de Vries P, Magnee P and Klapwijk T M 1992 \PRL \textbf{69} 510

\bibitem{Arutyunov}
Arutyunov K Yu, Auraneva H-P and Vasenko A S 2011 \PR B \textbf{83} 104509

\bibitem{Brinkman2003}
Brinkman A, Golubov A A, Rogalla H, Wilhelm F K and Kupriyanov M Yu 2003
\PR B \textbf{68} 224513

\bibitem{Brinkman2000}
Brinkman A and Golubov A A 2000 \PR B \textbf{61} 11297

\bibitem{Kupriyanov1999}
Kupriyanov M Yu, Brinkman A, Golubov A A, Siegel M and Rogalla H 1999 {\it Physica} C \textbf{326-327} 16

\bibitem{Heslinga1993}
Heslinga D R and Klapwijk T M 1993 \PR B \textbf{47} 5157

\bibitem{Nazarov}
Nazarov Yu V 1999 {\it Superlatt. Microstruct.} {\bf 25} 1221

\end{thebibliography}
\end{document}